%% file: antpaper_arxiv.tex
\documentclass[pre,twocolumn,superscriptaddress,showpacs,longbibliography]{revtex4-2} % Revtex class for arXiv submission

\usepackage{amsmath,amssymb,amsfonts}
\usepackage{graphicx}
\usepackage{hyperref}
\usepackage{float}
\usepackage{color}

\newcommand{\dd}{\mathrm{d}}
\newcommand{\X}{\mathbf{X}}

\newcommand{\e}{\mathbf{e}}

\usepackage{amsfonts}
\usepackage{amssymb}
\usepackage{amsmath}
\usepackage{graphicx}

\newcommand{\refSI}{Appendix}
\newcommand{\refMM}{Methods}

\DeclareMathOperator{\sgn}{sgn}

\usepackage{xr}

\begin{document}

\title{Argentine ants regulate traffic flow with stopped individuals}

\author{Ulrich Dobramysl}
\affiliation{Department of Applied Mathematics and Theoretical Physics, University of Cambridge, Cambridge CB3 0WA, United Kingdom}
\affiliation{Peter Medawar Building for Pathogen Research, University of Oxford, Oxford, United Kingdom}
\author{Simon Garnier}
\affiliation{Department of Biological Sciences, New Jersey Institute of Technology, Newark, NJ 07102, USA}
\author{Laure-Anne Poissonnier}
\affiliation{Research Center on Animal Cognition (CRCA), Center for Integrative Biology (CBI), Toulouse University, CNRS, UPS, 31062 Toulouse, France}
\affiliation{Animal Comparative Economics Laboratory, Chair of Zoology and Evolutionary Biology, University of Regensburg, Regensburg, Germany}
\author{Audrey Dussutour}
\affiliation{Research Center on Animal Cognition (CRCA), Center for Integrative Biology (CBI), Toulouse University, CNRS, UPS, 31062 Toulouse, France}
\author{Maria Bruna}
\affiliation{Department of Applied Mathematics and Theoretical Physics, University of Cambridge, Cambridge CB3 0WA, United Kingdom}
\affiliation{Mathematical Institute, University of Oxford, Oxford OX2 6GG, UK}

\begin{abstract}
We investigated the emerging traffic patterns of Argentine ants (\textit{Linepithema humile}) as they navigated a narrow bridge between their nest and a food source. By tracking ant movements in experiments with varying bridge widths and colony sizes and analyzing the resulting trajectories, we discovered that a small subset of ants stopped for long periods of time, acting as obstacles and affecting traffic flow. Interestingly, the fraction of these stopped ants increased with wider bridges, suggesting a mechanism to reduce traffic flow to a narrower section of the bridge.
To quantify transport efficiency, we measured the average speed of the ants on the bridge as a function of the pressure of ants arriving at the bridge, finding this relationship to be an increasing but saturating function of the pressure. We developed an agent-based model for ant movement and interactions to better understand these dynamics. Including stopped agents in the model was crucial to explaining the experimental observations.
We further validated our hypothesis by introducing artificial obstacles on the bridges and found that our simulations accurately mirrored the experimental data when these obstacles were included. These findings provide new insights into how Argentine ants self-organize to manage traffic, highlighting a unique form of dynamic obstruction that enhances traffic flow in high-density conditions. This study advances our understanding of self-regulation in biological traffic systems and suggests potential applications for managing human traffic in congested environments.
\end{abstract}

\maketitle

\section{Introduction}
Traffic, the movement of objects along bounded routes, is an integral part of all biological systems. It is involved, for instance, in the movement of transport vesicles inside cells \cite{mellman2000road,van2022challenges}, the distribution of nutrients and respiratory gases through the circulatory system \cite{tu2012computational}, and the transfer of information through the nervous system \cite{seguin2018navigation}. At the superorganism level, traffic is also critical to the functioning of social systems, from exchanging materials and goods through transportation networks to transferring information through communication networks \cite{coe2015global}.

Due to its central role in living systems, traffic is often tightly regulated by mechanisms that have evolved to maintain its stability and efficiency. When these regulation mechanisms fail, the consequences for the affected individuals can be disastrous: blood clots can lead to strokes and heart attacks, deregulation of brain electric activity can lead to epileptic episodes \cite{bonansco2016plasticity}, and mass panic in crowded environments can lead to deadly stampedes \cite{shiwakoti2011animal}.

In many cases, biological traffic is regulated by well-defined regulatory circuits: sensors measure proxies of traffic activity and send feedback to effectors that can act to increase or decrease its intensity or modify its characteristics \cite{chowdhury2005physics}. In other cases, however, traffic regulation mechanisms are not explicit. Instead, they emerge from the interactions between the objects that constitute the traffic itself \cite{dussutour2004optimal,czaczkes2013negative,wendt2020negative}. Pedestrian traffic, most notably in humans and ants, and vehicular traffic are prime examples of self-organizing traffic whose stability and efficiency depend primarily on the behavior of the individuals that compose it \cite{Moussaid.2011,moussaid2009collective,dussutour2004optimal}. In the absence of external regulatory circuits, the rules individuals follow when moving with and around each other determine the organization of the traffic. When these rules are not adequately adjusted to the traffic density, undesirable phenomena such as phantom traffic jams, turbulences, and ultimately complete standstill will appear \cite{helbing2001traffic,duan2023spatiotemporal}.

In this context, some species of ants have shown incredible resiliency in their traffic, even under extreme crowdedness \cite{burd2002traffic, couzin2003self, dussutour2004optimal, fourcassie2010ant, gravish2015glass, honicke2015effect, strombom2018self, Poissonnier2019}. In typical examples of self-organizing traffic, traffic flow increases with traffic density up to a critical density beyond which overcrowding takes effect and the relationship becomes inverted \cite{prigogine1961boltzmann, helbing2001traffic}. However, a previous study in the Argentine ant \textit{Linepithema humile} showed that the ants could maintain a constant and high-intensity traffic flow well beyond this critical density \cite{Poissonnier2019}. This feat was explained by the ants adjusting their walking speed and limiting their interactions as the density of neighboring ants grew, thereby mitigating the negative consequences of the increased interaction rate.

Here, we analyzed these experiments using an advanced data extraction method.
In particular, we generated an exhaustive catalog of 2.6 million trajectories over a wide range of traffic densities, allowing us to analyze the behavior of Argentine ants under crowded conditions with unprecedented detail. We found that even at high traffic flow, some ants stop in the middle of the traffic and remain motionless for long periods, seemingly obstructing parts of the available route. We evaluated the impact of these stopped ants on the traffic organization using an agent-based model whose hypotheses are formed directly from the analysis of the ant trajectories. Contrary to intuition, the presence of stopped ants in the simulations improved traffic flow and contributed significantly to the ability of Argentine ants to maintain traffic flow beyond the critical density. We validated these findings with further experiments in which artificial obstacles were placed along the route to mimic the presence of stopped ants.

This result is reminiscent of previous work on human pedestrian traffic in which the presence of obstacles in the urban environment may help facilitate traffic in congested areas \cite{luo2018experimental, helbing2002simulation,araujo2023steering}. However, these obstacles are placed there by external agents and are not the product of the pedestrians' activities. To our knowledge, it is the first time that this form of traffic management strategy has been described, where the participants use dynamic obstruction to improve traffic flow. Our results could inspire new approaches for reducing congestion, such as using autonomous vehicles as adaptive roadblocks to facilitate pedestrian and vehicular traffic at high density.

\begin{figure}[tb]
\centering
\includegraphics[width=\linewidth]{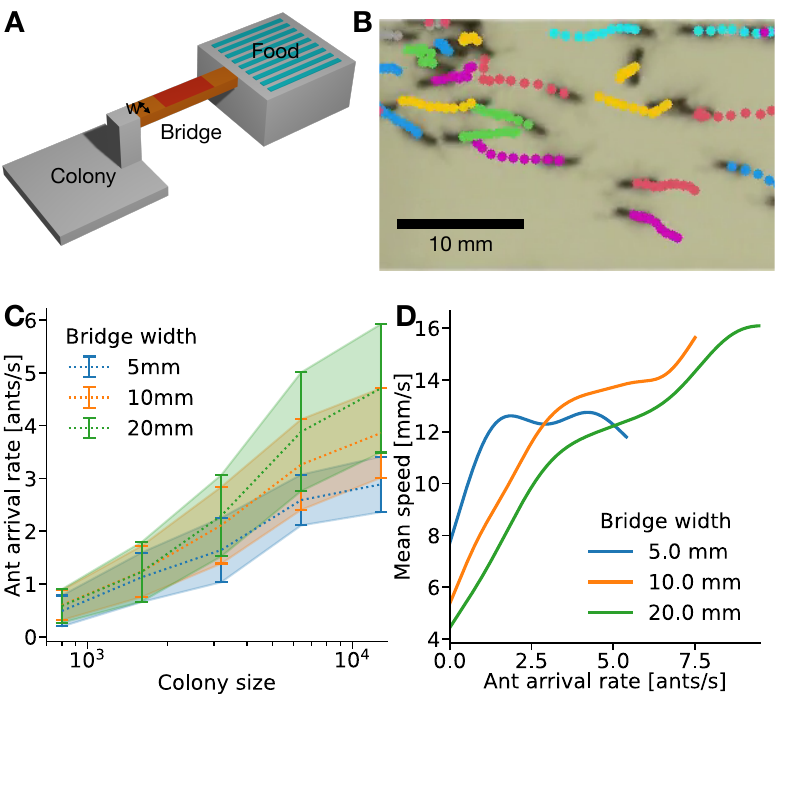}
\caption{{\bfseries Ant traffic across a narrow bridge.} (A) Schematic of the experimental setup: Ants enter a narrow bridge (width $w =\,$5, 10, or 20 mm) from the colony on the left, traverse to the food source, and return to the colony. Tracking of ants occurs in the red area. (B) Tracking video frame from an experiment with a bridge of $20$ mm width and a colony of approximately 6\,400 individuals. Different colors indicate different ant tracks, with the points representing the ant positions throughout the last ten frames. (C) Arrival rate of ants at either end of the bridge as a function of the colony size. Solid lines show the arrival rate averaged over time (after the ant traffic has reached a steady state, Fig.~\ref{fig:figureS1}A) and all experiments with a given colony size (Table~S1), shaded regions indicate the standard deviation. (D) Average ant speed as a function of the arrival rate.} 
\label{fig:figure1}
\end{figure}

\section{Results}
Here, we investigated how ants successfully maintain a smooth traffic flow and avoid traffic congestion under the broadest possible range of densities. We used Argentine ants \textit{Linepithema humile} from the European super-colony. In the experiments, colonies starved for five days were connected to a sugar solution (1M) for an hour using a plastic bridge (Fig.~\ref{fig:figure1}A). A combination of bridge widths (5, 10, and 20 mm) and colony sizes (400 to 12\,800 ants) were used to manipulate density. A total of 116 videos were analyzed starting when a bidirectional steady flow was observed. 

\begin{figure*}%[tbhp]
\centering
\includegraphics[width=\linewidth]{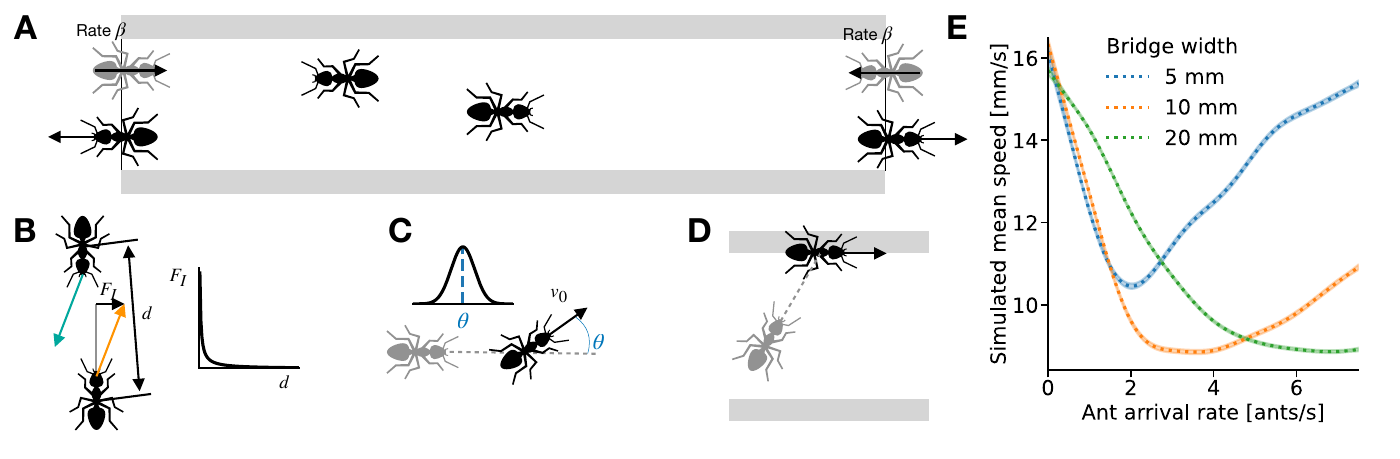}
\caption{{\bfseries Agent-based stochastic model for ant movement.} (A) Agents (simulated ants) enter the bridge from either side and are removed once they return to either boundary. (B) Agents interact with each other using a force $F_I$ that decreases exponentially with distance $d$. (C) Agents move at a constant speed $v_0$ and change their direction randomly according to a rotational diffusion process.  (D) When reaching the edge of a bridge, agents align their direction with the edge. (E) Average ant speed as a function of the arrival rate from simulations. Note the high speeds at low arrival rates, a feature not present in the experimental data (cf. Fig.~\ref{fig:figure1}D).}
\label{fig:figure2}
\end{figure*}

\subsection{Ant arrival rate increases with colony size}

We first looked at the arrival rate of ants at the bridge from both the colony and the food source over time. To this end, we counted the number of ants arriving at either end in $10$ second intervals over the whole duration of the experiments. We did this separately for each experiment and averaged the resulting ant arrival rate against time data over experiments with the same colony size, shifting time such that the first ant to arrive at the bridge did so at time zero. The resulting curves show a steep increase of ants entering the bridge from the nest before traffic settles into a steady state (Fig.~\ref{fig:figureS1}A). The buildup rate of this initial increase strongly depends on the colony size, with the largest colony having the biggest onrush of ants (Fig.~\ref{fig:figureS1}A). Traffic stemming from ants arriving from the food source is delayed by roughly four minutes owing to the exploration of the bridge and the food source before returning to the bridge (Fig.~\ref{fig:figureS1}A). Once the traffic along the bridge reaches a steady state (after roughly 1\,000 s), the arrival rates from the nest and the food source show no appreciable difference (Fig.~\ref{fig:figureS2}). Steady-state ant arrival rates increase monotonously with the colony size (Fig.~\ref{fig:figure1}C). For the smallest colony sizes, the ant arrival rate shows hardly any dependence on the width of the bridge; however, for larger colony sizes, this rate starts to deviate between the three bridge sizes (Fig.~\ref{fig:figure1}C, Fig.~\ref{fig:figureS1}B). This indicates a saturation of the bridge entry points for larger colonies, which could have more individuals entering the bridge were it to be wider. We conclude that the ant arrival rate is a more precise measure for traffic pressure than the colony size, and consequently we will subsequently use the arrival rate in our analysis. 

\subsection{Mean ant speed increases with ant arrival rate}
To find out how traffic pressure affects the speed of the ants, we next divide each experiment into 10 s time intervals. We measure the rate at which the ants arrive at the bridge and the mean speed of the ants on the bridge during each time interval. 
This allowed us to calculate the average ant speed as a function of the ant arrival rate, or traffic pressure, at the bridge by fitting a generalized additive model (GAM) to the data (Fig.~\ref{fig:figure1}D, \refMM). The average ant speed increases monotonically with the pressure, saturating at between $12$-$16$ mm/s at high ant arrival rates. The saturation scales with the bridge width, occurring at lower pressures for smaller bridges. For the 10mm and 20mm bridges, this saturation point occurs at an arrival rate of just over $2.5$ ants/s, while for the 5mm bridge, it occurs at around one ant/s. The ant speed ranges between 12 and 16 mm/s during this later phase.

\subsection{Agent-based model for ant traffic}
To understand the speed against traffic pressure relationship evident in Fig.~\ref{fig:figure1}D, we turned to an agent-based stochastic model for ant movement (see \refMM{} for details). Here, ants are released randomly at either end of a bridge and are removed from the simulation once they leave the bridge (Fig.~\ref{fig:figure2}A). They change direction in small, random increments (Fig.~\ref{fig:figure2}B) according to a rotational diffusion process. They move at constant speed unless they come close to other ants, whereupon they change speed and direction according to an interaction force that decays exponentially with distance (Fig.~\ref{fig:figure2}C). From the bridge experiments, we also verified that ants prefer moving along a boundary rather than being repelled by it. We implemented this in our model by having an ant align its direction with the boundary when it hits one (Fig.~\ref{fig:figure2}D). We extracted the agent-based model's parameters from our experimental data (see Methods section for details) and simulated the experimental setup as closely as possible.

From our simulation data, we generated a graph showing the average speed on the bridge as a function of the simulated ant arrival rate (Fig.~\ref{fig:figure2}E). For ant arrival rates greater than 3 s$^{-1}$, this graph displays the same monotonic increase of the average ant speed as the experimental results show (compare with Fig.~\ref{fig:figure1}D). In contrast, the simulation data show a comparatively high average speed for small ant arrival rates. This means that at low numbers of ants and low external pressure, ant traffic is essentially unhindered. This feature is absent in the experimental results; hence, the present agent-based model fails to explain the experimental data. 

\subsection{Stopped ants play an important role in shaping traffic}
\begin{figure}[t]
\centering
\includegraphics[width=\linewidth]{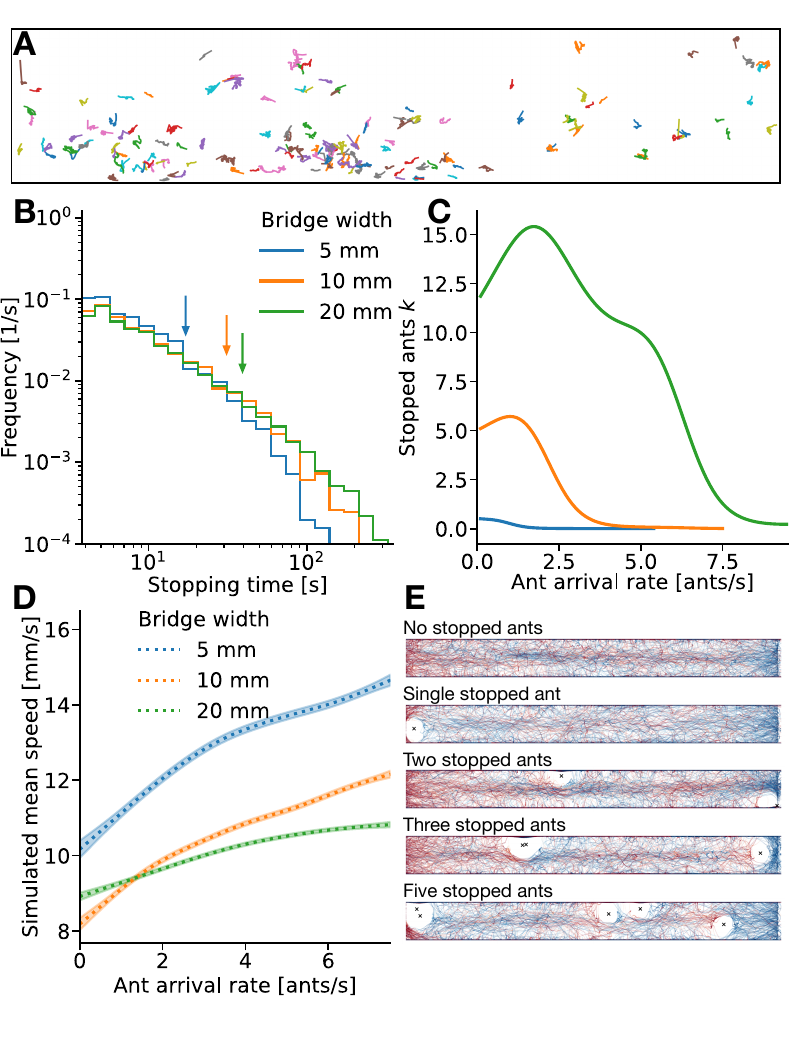}
\caption{{\bfseries Ant speed depends on ant influx and stopped individuals.} (A) Positions of stopped ants over a 10-minute interval. (B) Histogram showing the time ants stop for. Arrows indicate the average stopping time. (B) Average number of ants stopped as a function of the ant arrival rate. (D) Average ant speed versus ant arrival rate from individual-based simulations when stopped ants are present (three, five, and ten stopped ants for the 5, 10, and 20 mm bridge, respectively). (C) Simulated ant tracks (red tracks start at the left end, while blue tracks start at the right end of the bridge) with 0, 1, 2, and 5 (top to bottom) obstacles inserted at random points on the bridge for a constant ant arrival rate of 5 ants/s.}
\label{fig:figure3}
\end{figure}

The experiments indicate that a fraction of ants randomly stop and become stationary for significant periods of time. The tracks shown in Fig.~\ref{fig:figure3}A highlight that a small subset of ants stays in place, seemingly acting as obstacles to other ants. Most stopped ants are outside the area covered by regular ant traffic, and the spatial distribution of stopping positions is non-uniform (\refSI, Fig.~\ref{fig:figureS5}). The average time stopped ants spend motionless is between $15$-$40$ seconds depending on the width of the bridge (Fig.~\ref{fig:figure3}B). The average number of stopped ants depends again on the current ant arrival rate, with higher arrival rates leading to fewer stopped ants (Fig.~\ref{fig:figure3}C). Crucially, the narrower the bridge, the smaller the number of stopped ants. 

From this data, we hypothesized that stopped ants actively shape traffic into narrower channels by artificially reducing the available space. In our simulation model, we implemented different numbers of stopped ants at random positions on the bridge to understand how their traffic influences the transport across the bridge. Simulated ant trajectories avoided the stopped ants, creating space devoid of traffic (Fig.~\ref{fig:figure3}E). Indeed, stopped ants cause the simulated mean speed across the bridge to drop at low ant arrival rates. As a result, the model incorporating stopped ants captures the experimental data well (Fig.~\ref{fig:figure3}E).

\subsection{Artificial obstacles play a similar role to stopped ants}
\begin{figure}[tbhp]
\centering
\includegraphics[width=\linewidth]{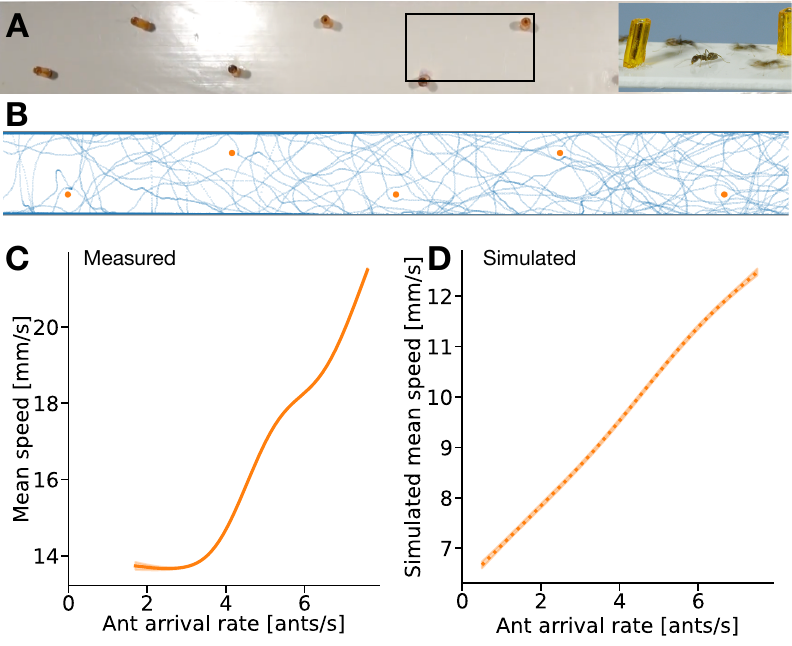}
\caption{{\bfseries Artificial obstacles affect the traffic flow similarly to stopped individuals.} (A) Experimental setup of a bridge of 10 mm with pillars. (B) Simulation showing pillars in orange and ant tracks over 100 s in blue. 
(C) Average ant speed as a function of ant arrival rate (measured over $10$ s time intervals) for bridges with artificial obstacles. (D) Simulated average ant speed as a function of ant arrival rate with artificial obstacles.}
\label{fig:figure4}
\end{figure}

To test this further, we introduced artificial obstacles (pillars) in a regular pattern across the 10 mm bridge (Fig.~\ref{fig:figure4}A) connected to a colony of approximately 25\,600 individuals. This caused ants at low arrival rates (up to 3 ants/s) to move substantially faster (13.5 mm/s, up to twice the speed) than in the same bridge without obstacles (compare Fig.~\ref{fig:figure1}D and Fig.~\ref{fig:figure4}C). At higher arrival rates, we observe a sharp increase in the mean speed up to 20 mm/s (Fig.~\ref{fig:figure4}C). In all, the pillars reduce the space available to ants, leading to an increased speed and shifting the curve in Fig.~\ref{fig:figure1}D so that the plateau is attained at lower ant arrival rates (we cannot access an analogous sharp increase beyond 14 mm/s in the case without obstacles, which might occur for ant arrival rates higher than 8 ants/s). In the presence of pillars, we found virtually no stopped ants in the bridge (Fig.~\ref{fig:figureS3}). This supports our hypothesis that pillars can replace the role of stopped ants in actively shaping traffic by reducing the available space. To validate it, we implemented pillars in our model arranged in the same regular pattern as in the experiments (Fig.~\ref{fig:figure4}B). We find that the simulated mean speed describes the experimental data with pillars well (Fig.~\ref{fig:figure4}D). Comparing the results with pillars or stopped ants, we find that in experiments, the regular arrangement of pillars produces larger speeds than the corresponding experiment of dynamically stopped ants (\refSI, Fig.~\ref{fig:figureS6}A). In the case of simulations, the same is true only at high densities, with the effect reversed at lower densities (\refSI, Fig.~\ref{fig:figureS6}B). 

\section{Discussion}

In this paper, we demonstrated that ants autonomously regulate their own traffic and effectively overcome environmental constraints. Our findings reveal a remarkable efficiency in ant transport, particularly under high-density conditions, where typical traffic systems, such as pedestrian or vehicular models, would predict a decline in speed. Contrary to these systems, we observed that not only does the flux (or transport rate) remain high at elevated densities, but the velocity of individual ants also reaches its maximum in high-traffic situations (Fig.~\ref{fig:figure1}). This result challenges conventional traffic dynamics, where increased density generally leads to slower movement, even if the overall flow is maintained.

We developed an agent-based model simulating ant movement along the bridge to interpret these observations further. This model allowed us to examine the underlying mechanisms driving this unique traffic behavior. Our analysis suggests that when traffic density is moderate, ants coordinate their movements by having a fraction of individuals stop. These stationary ants reduce the effective cross-sectional area of the bridge, thereby concentrating the movement into a narrower, more efficient region. This counterintuitive behavior--where stopped ants enhance overall flow--was supported by our simulations, which revealed that the presence of motionless ants is crucial for replicating the experimental results.

To validate our hypothesis, we introduced artificial obstacles in both experiments and simulations to mimic the effect of stopped ants. With these regularly spaced pillars in place, we observed fewer stationary ants, yet the traffic flow remained efficient. This suggests that the artificial obstacles replicate the function of the stopped ants by inducing a higher density in localized areas, thus reducing the need for ants to halt. Our findings confirm that these obstacles optimize traffic flow in a manner similar to the natural stoppage behavior of the ants.

These results offer a novel perspective on traffic regulation strategies in biological systems, where crowding can paradoxically lead to improved transport efficiency. In the case of Argentine ants, this improvement seems to arise from enhanced coordination and more directed movement under crowded conditions. This form of dynamic, self-organized traffic management, where individuals adjust their behavior to maximize collective efficiency, could have implications beyond ant colonies. Our findings suggest potential applications in human traffic systems, particularly in crowded environments such as stadiums or transit hubs. By strategically incorporating obstacles or barriers, achieving optimal flow while maintaining safety may be possible \cite{frank2011room,yano2018effect,hu2023experimental,zhao2020experimental}, offering new insights into the design of more efficient pedestrian traffic systems \cite{shiwakoti2013enhancing}. Studies on pedestrians indicate that the spatial layout of obstacles is important \cite{chen2019effect}. While our focus was on the effect of stopped individuals on the flow, we also observed that the spatial arrangement of such individuals can play a role.

\section*{Methods}\label{sec:methods}
\subsection*{Experimental protocol}
The experimental data used for the analyses in (sections without pillars) was taken from \cite{Poissonnier2019}. Here, we briefly summarise their experimental protocol; full details can be found in \cite{Poissonnier2019}. In the experiments, a colony of Argentine ants {\it Linepithema humile} was connected to a food source using a bridge of 170 mm in length. To manipulate density, the authors used a combination of bridges of different widths (5, 10, and 20mm) and experimental colonies of various sizes (from 400 to 25\,600 ants) (see Fig.~\ref{fig:figure1}A, \cite{Poissonnier2019}). They conducted a total of 170 experiments.  The size and design of the food source were such that it never saturated, even for the largest colony size.

In this paper, we introduced a new experiment conducted with the 10 mm and the 20 mm wide bridges similar to the one used by \cite{Poissonnier2019}. We glued eight pillars on the bridge, materialized by cylindrical plastic beads (4mm high, 1mm wide). The pillars were 2cm apart and placed in a staggered pattern (see Fig.~\ref{fig:figure4}). We run four replicates of this experiment following the same protocol as in \cite{Poissonnier2019}. Before each replicate, the experimental colonies were starved for five days, and the experiment started when the ants were given access to a food source placed on a platform (120x120 mm) at the other end of a plastic bridge.  The total length of the bridge was 170mm. The food consisted of a 1M sucrose solution contained in 16 grooves (185mm long) carved in a block of plexiglas. The grooves were numerous enough to prevent crowding effects at the food source. The whole experimental set-up was isolated from any sources of disturbance by surrounding it with white paper walls. 

\subsection*{Data collection, extraction, and analysis}
Throughout the experiments, the traffic on the bridge was filmed from above for 60 minutes, starting as soon as the first ant entered the bridge. We used two cameras (Canon LEGRIA HF G30): one was filming the entire set-up, and the second one was focused on the bridge. 
We analyzed 116 videos among the 170 available, using trackR~\cite{trackR} to extract the trajectories of individual ants~ (see Fig.~\ref{fig:figure1}B). We tracked a central portion of the bridge with a length of 100 mm. In Figs. \ref{fig:figure2}, \ref{fig:figure3}, and  \ref{fig:figure4}, we only use the data once a steady bidirectional flux along the bridge has been attained. 

To calculate the relationship between the mean ant speed and the ant arrival rate, we divided each experiment into 10 s intervals and counted how many ants arrived at the bridge. For each interval, we averaged the individual speeds of all ants on the bridge throughout that interval. We then fitted a GAM to the resulting mean ant speed versus ant arrival rate data using the \texttt{mgcv} package \cite{mgcv} in R (version 4.4.1), using six knots, restricted maximum likelihood as the smoothing parameter estimation method, and a Gaussian distribution with the identity link function (Figs.~\ref{fig:figure1}D and~\ref{fig:figure4}C).

To identify stopped ants, we looked for parts of individual ant tracks that did not move more than 1 mm in any given 4 s interval. We gathered information on how long ants stay stopped according to these criteria (Fig.~\ref{fig:figure3}B). We also counted how many ants were stopped on average throughout each 10 s time window and again fitted a GAM to the resulting stopped ants vs ant arrival rate data, using 6 knots and a Poisson distribution with a logarithmic link function (Fig.~\ref{fig:figure3}C).

\subsection*{Agent-based model}
Ants are modeled as interacting active Brownian particles, where the $i$th ant with center $\X_i(t) = (X_i(t), Y_i(t))$ and orientation $\Theta_i(t)$ at time $t$ evolves according to 
\begin{subequations}
	\label{abm_model}
\begin{align}
\label{sde_x}
	\dd \X_i &= v_0 \e(\Theta_i)\dd t - \sum_{j\ne i} \nabla u(|\X_i-\X_j|) \dd t,\\
	\label{sde_theta}
    \dd \Theta_i &= \sqrt{2D}\dd W_i.
\end{align}
\end{subequations}
Here $v_0$ is the (fixed) self-propulsion speed, $\e(\theta) = (\cos \theta, \sin \theta)$ is the ant direction, $D$ is the rotational diffusion coefficient, and $W_i$ are independent one-dimensional Brownian motions. The ants interact through a repulsive exponential potential $u(r) = C_r \exp(-r/l_r)$, where $C_r$ and $l_r$ are the interaction's strength and range, respectively. Pheromone deposition and trails are not included in \eqref{abm_model} since its effect is negligible given the experiment bridge geometry and level of crowding. 

We simulate the tracked central portion of the bridge, that is, the domain  $(x, y) \in [-50, 50] \times [-w/2, w/2]$. Ants are released at $x = \pm 50$ mm uniformly distributed in $y$ and with orientations $\cos(\theta) = \pm 1$, respectively. The boundaries $x = \pm 50$ mm are absorbing (when an ant arrives there, it is removed from the simulation). In experiments, some ants can walk sideways along the bridge ends $y = \pm w/2$. In the simulations, we model it as a slip boundary: if an ant hits $y = \pm w/2$, we project its direction $\e(\Theta)$ along the bridge, becoming $\e(\Theta) = (\sgn(\cos(\Theta)), 0)$  (see Fig.~\ref{fig:figure2}D).  In the simulations with pillars (Fig.~\ref{fig:figure4}D), the interaction between an ant and a pillar is modeled with the same interaction potential $u$ as the ant-ant interaction. 
Trajectories of \eqref{abm_model} are simulated with an Euler--Maruyama scheme with $\Delta t = 0.1$ s using \texttt{Agents.jl}. The parameter values, obtained by fitting to the experimental data, are $v_0 = 15.5$ mm/s, $D_R = 0.38$ rad$^2$/s, $C_r=129$ mm$^2$/s and $l_r=3.96$ mm (see \refSI{} for details).

In simulations with $k$ stopped ants, we choose $k$ random positions on the bridge at the start of the simulation run and placed ants held motionless at those positions throughout the simulation at these positions.

For each value of the ant arrival rate, we ran 200 realizations of 100 seconds each. We ascertained that the ant traffic reached a steady state after at most 50 s and used the latter 50 s of each realization to calculate the average speed of ants. We fitted the same GAM model to the simulated data as for the experiments, except for the number of knots (we used ten instead of 6 knots, Figs.~\ref{fig:figure2}E, \ref{fig:figure3}D and \ref{fig:figure4}D).

\begin{acknowledgements}
UD acknowledges funding from the Isaac Newton Trust (G101121).
SG was supported by the National Science Foundation under grant no \#EF-2222418. 
	MB acknowledges funding from the Royal Society (URF$\backslash$R1$\backslash$180040).
\end{acknowledgements}

% Bibliography

\onecolumngrid
\break

\appendix
\section*{Appendix}
\renewcommand{\thefigure}{S\arabic{figure}}
\renewcommand{\thetable}{S\arabic{table}}
\setcounter{figure}{0}
\setcounter{table}{0}

\subsection*{Agent-based model fitting}

\begin{figure}[thb]
\centering
\includegraphics[height=0.3\linewidth]{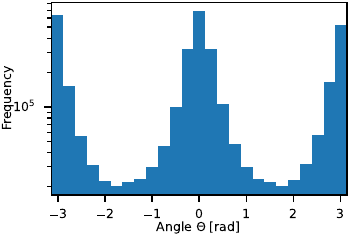} \qquad \includegraphics[height=0.3\linewidth]{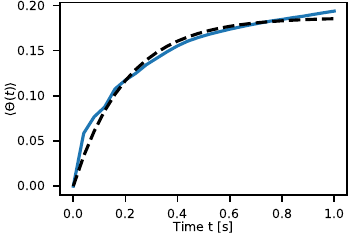}
\caption{{\bfseries Rotational diffusion coefficient fitting procedure.} (A) Histogram of the orientations $\Theta(t_k)$ of ants approximated as the angle $\angle[\X(t_k+\Delta t) - \X(t_k)]$. (B) Mean-square angular deviation $\langle \Theta^2(t) \rangle$ from data (blue line) and theoretical fit \eqref{msd_t} (black dash line).}
\label{fig:DR}
\end{figure}

The agent-based model (ABM) \eqref{abm_model} contains four parameters we fit using the experimental data. The rotational diffusion coefficient $D$ are fitted using isolated ant trajectories $\X_i(t)$ in low crowding conditions, such that the interaction term in \eqref{sde_x} can be neglected. An isolated trajectory is such that no other ants are within a circle of 9 mm of radius. More precisely, isolated trajectories are
$$
\{\X_i(t), T_1\le t\le T_2,  \text{ such that }  \|\X_i(t)- \X_j(t) \| > 9\text{ mm for all } j\ne i \text{ and } t \in [T_1, T_2]\}.
$$  
We use the mean-square angular deviation in time $t_j$ to estimate the rotational diffusion $D$. To do so, we first need to account for the effect of the bridge boundaries at $y = \pm w/2$ on the angle. They act as a confining potential in angle, making ants prefer to travel along the bridge (see Fig.~\ref{fig:DR}A). The distribution in angles is well approximated by an Ornstein--Uhlenbeck process
\begin{equation}
	 \dd \Theta_i = -k \Theta_i \dd t + \sqrt{2D}\dd W_i,
\end{equation}
which satisfies
\begin{equation}
\label{msd_t}
	\langle \Theta^2(t) \rangle = \frac{D}{k} \left( 1- e^{-2 k t} \right). 
\end{equation}
We then used least-squares fitting of the mean-square angular deviations of isolated trajectories to \eqref{msd_t} (see Fig.~\ref{fig:DR}B).

The procedure to fit the self-propulsion speed $v_0$ and the strength and the range of the interaction potential $u(r) = C_r \exp(-r/l_r)$ is as 
follows. Given the pairwise and short-range nature of the interaction, we use pairs of trajectories of ants that approach each other from opposite directions and with any other ants sufficiently far away. Specifically, we identify pairs of ants $\{i,j\}$ such that $\|\X_i(t)-\X_j(t)\|< 1.75$ mm at some time $t$ and $\|\X_i(t)-\X_k(t)\|> 5$ mm for all $k \ne i,j$. Fixing $i = 1$ and $j = 2$, it means \eqref{sde_x} reduces to
\begin{equation*}
	\dd \X_1 = v_0 \e(\Theta_1)\dd t + {\bf F} \dd t,\qquad 
		\dd \X_2 = v_0 \e(\Theta_2)\dd t - {\bf F} \dd t,
\end{equation*}
where ${\bf F} = - u'(r) \X/ r$ is the pairwise force, $ \X = \X_1-\X_2$ is the relative position and $r = \|\X\|$, or
\begin{equation}
	\dot \X = v_0 (\e(\Theta_1) - \e(\Theta_2)) + 2 {\bf F} .
\end{equation}
Rearranging and taking absolute values, we arrive at
\begin{align}\label{squared}
	|\dot \X - 2 {\bf F}|^2 = \dot r^2 + (2 u')^2 + 4 \frac{u'}{r} \dot \X \cdot \X = v_0^2 |\e(\Theta_1) - \e(\Theta_2)|^2 = 2 v_0^2 [1 - \e(\Theta_1)\cdot \e(\Theta_2)].
\end{align}

\begin{figure}[htb] % h = here, t = top, b = bottom, p = page of figures
    \begin{center}
    	    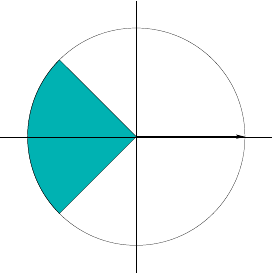
    \end{center}
    \caption{Region of `head-on' interactions. If, say, $\e(\Theta_1) = (1,0)$ (fixed), then $\e(\Theta_2)$ averaged over time is allowed to be in the shaded region, $\Theta_2 \in [3\pi/4, 5\pi/4]$.}
    \label{fig:sketch_angle}
\end{figure}

\begin{figure}[tbh]
\centering
\includegraphics[width=.6\linewidth]{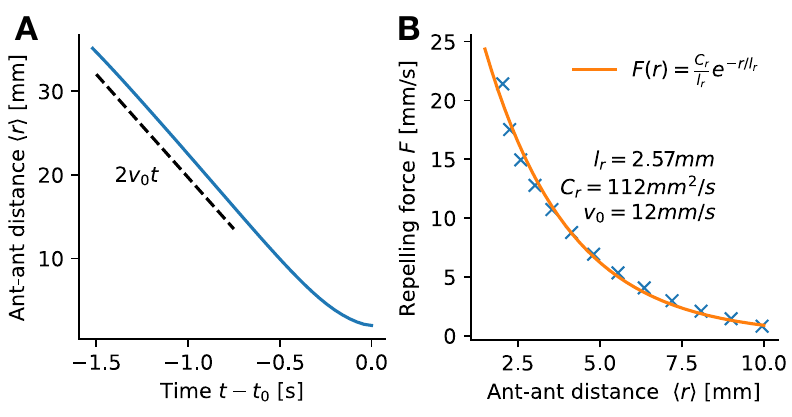}
\caption{{\bfseries Extraction of the ant-ant interaction parameters.} (A) Ant-ant pair distance averaged over all identified 'head on' interaction pairs. Time is measured relative to the pair's closest approach time $t_0$. The average ant speed $v_0$ is given as the twice the slope of the distance curve when $t-t_0 < 0.75$s. (B) The empirical force (blue crosses) with the Morse force expression fitted. The extracted parameters of the ant-ant interaction force are given in the graph.} 
\label{fig:figureS7}
\end{figure}

We denote by $t_0$ the time of closest approach. Out of the isolated pairs of trajectories, we select `head-on' interactions or pairs of trajectories whose relative orientation, averaged over an interval of 80 frames or 3.2s centered at $t_0$, lies within $[3\pi/4,5\pi/4]$ (see Fig.~\ref{fig:sketch_angle}). 
Otherwise, the interaction pair is not considered head-on and is rejected for this analysis. 
We then extract the relative distance between the two ants before the collision as a function of time and average over trajectories $r(t-t_0)$ (so that the point of closest approach is the same for all pairs of trajectories). The result is shown in Fig.~\ref{fig:figureS7}(A). 
If the interaction is a perfect head-on interaction, then $\e(\Theta_1)\cdot \e(\Theta_2) = -1$ and $\X \cdot \X = \dot r r$ and  \eqref{squared} reduces to
\begin{align}\label{perfect_headon}
	\dot r^2 + (2 u')^2 + 4 u'r = (\dot r + 2 u')^2 = 4 v_0^2  \quad	\Longrightarrow \quad  \dot r = 2(v_0 - u').
\end{align}
Our selection of `head on' trajectories justifies using \eqref{perfect_headon} in our fitting procedure.\footnote{In fact, we also fitted the parameters using \eqref{squared}, and no noticeable difference to the resulting parameter values was found from the fitted values using \eqref{perfect_headon}.} Far away from $t_0$, $r(t-t_0)$ is very well approximated by a linear fit (Fig.~\ref{fig:figureS7}(A)), which indicates that the interaction is negligible at those distances (or that the range of the interaction $\ell_r$ is small compared to the relative distance $r(t-t_0)$ for large times $t_0-t$). Hence, we may extract $v_0$ from the slope of $r(t-t_0)$ far from $t_0$. We find that $v_0=12$mm/s. The final step is to fit the force magnitude, $F = -u'(r)$. Having ascertained that no other ants are close to the interacting pair, we can assume that the only force acting on the pair is the interaction force, and thus $\dot{r}(t)=2v_0+2 F(r(t))$. We calculate the instantaneous relative speed $\dot{r}(t_k)$ as the central 
difference of $r(t_k)$. Rearranging  this relationship gives $F(r)=\dot r/2 - v_0$ (Fig.~\ref{fig:figureS7}B).
We then fit the magnitude of the force $F(r)=C_r/l_r\exp(-r/l_r)$ using a least squares fitting procedure and arrive at the parameter values $C_r=112$mm$^2$/s and $l_r=2.57$mm.

\begin{figure}%[tbhp]
\centering
\includegraphics[width=.8\linewidth]{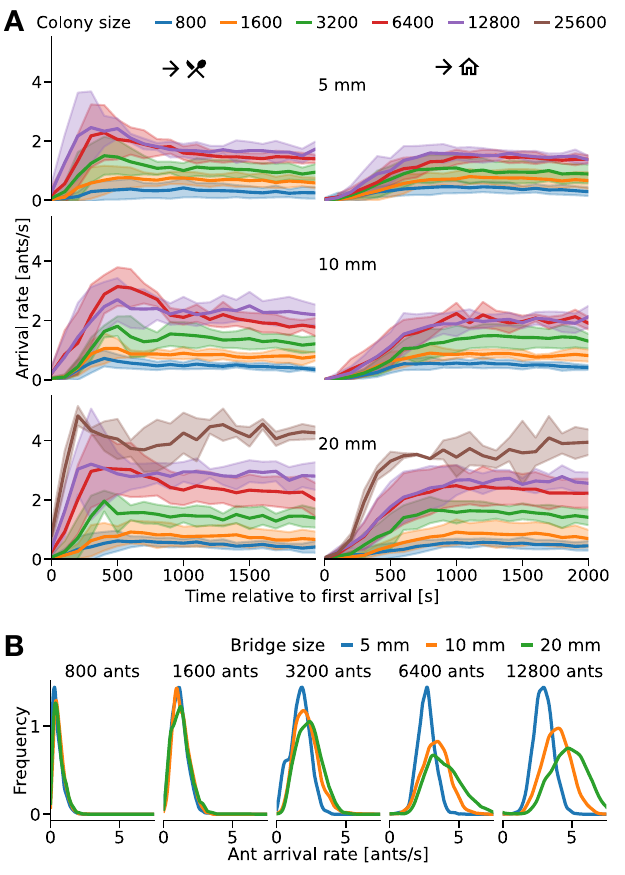}
\caption{{\bfseries Ant arrival rates from the nest and the food source} (A) Arrival rates of ants at the ends of the bridge closest to the nest (left column) or the food source (right column) for the 5mm, 10mm, and 20mm bridges (first, second and third row respectively) as a function of time. Time zero is set as the arrival of the first ant at the bridge. (B) Distribution of ant arrival rates at steady state ($t> 1000$ s) for different bridge widths and colony sizes. Plots show the kernel density estimation of the ant arrival rates.}
\label{fig:figureS1}
\end{figure}

\begin{figure}%[tbhp]
\centering
\includegraphics[width=.9\linewidth]{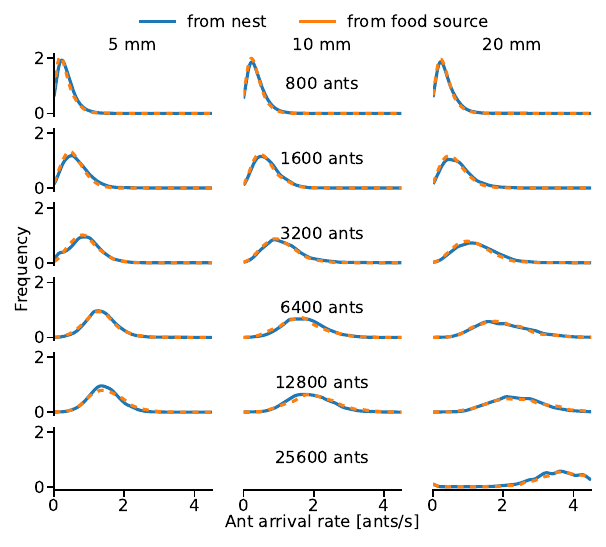}
\caption{{\bfseries Arrival rates from the nest and from the food source show similar distributions.} Distributions of ant arrival rates from the nest and the food source for the different bridge widths (columns) and colony sizes (rows). Plots show the kernel density estimation of the ant arrival rates with a bandwidth of $0.1\operatorname{s}^{-1}$.}
\label{fig:figureS2}
\end{figure}

\begin{figure}%[tbhp]
\centering
\includegraphics[width=.5\linewidth]{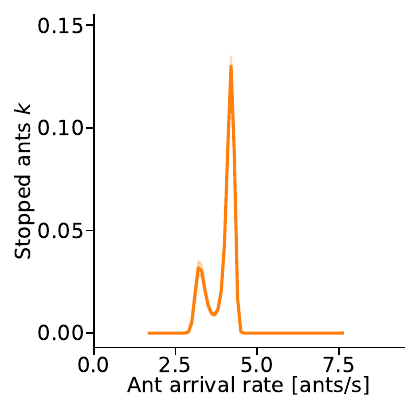}
\caption{{\bfseries Number of stopped ants in experiments with pillars.}}
\label{fig:figureS3}
\end{figure}

\begin{figure}%[tbhp]
\centering
\includegraphics[width=.9\linewidth]{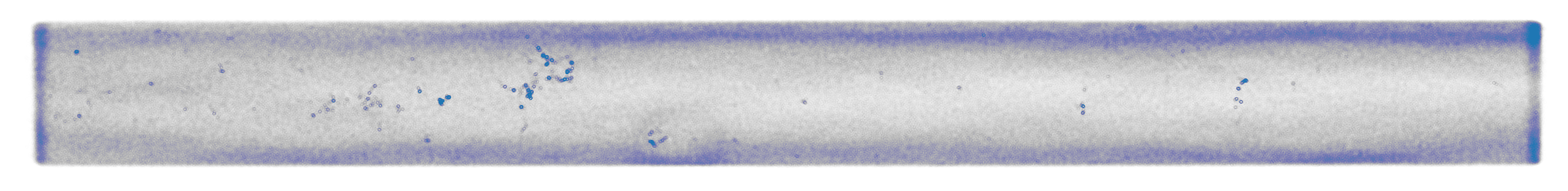}
\caption{{\bfseries Superposition of ant positions over a 10-minute interval.} It shows most stopped ants accumulate at the lower left part of the domain, with most traffic occurring along the top part. Fig.~\ref{fig:figure3}A shows only the stopped ants' positions in this experiment.  Experiment with the 20 mm bridge and 12,800 colony size. }
\label{fig:figureS5}
\end{figure}

\begin{figure}%[tbhp]
\centering
\includegraphics[width=.6\linewidth]{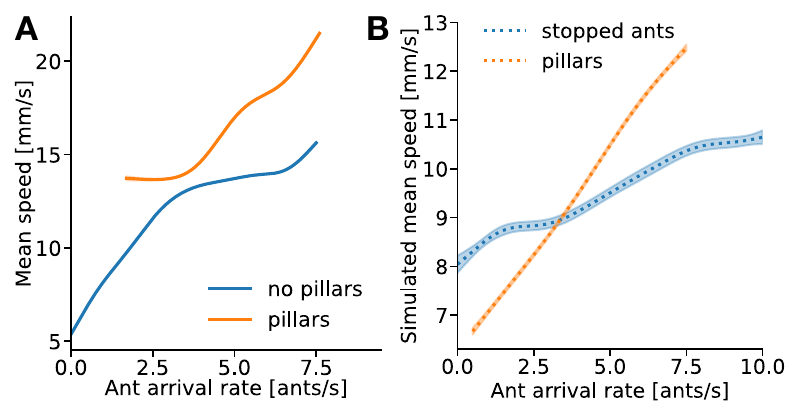}
\caption{{\bfseries Comparison between stopped ants and pillars.} (A) Mean speed vs ant arrival rate from experiments without (blue) and with pillars (orange). (B) Mean speed vs ant arrival rate from simulations with stopped ants (blue) and with pillars (orange).} 
\label{fig:figureS6}
\end{figure}

\end{document}

%% file: FigureS2.pdf_t
\begin{picture}(0,0)%
\includegraphics{FigureS2.pdf}%
\end{picture}%
%
%  encoding: UTF-8 
%
\setlength{\unitlength}{953sp}%
\begingroup\makeatletter\ifx\SetFigFont\undefined%
\gdef\SetFigFont#1#2#3#4#5{%
  \reset@font\fontsize{#1}{#2pt}%
  \fontfamily{#3}\fontseries{#4}\fontshape{#5}%
  \selectfont}%
\fi\endgroup%
\begin{picture}(9027,9024)(-14,-8173)
\put(  1,-2221){\makebox(0,0)[lb]{\smash{{\SetFigFont{5}{6.0}{\rmdefault}{\mddefault}{\updefault}{\color[rgb]{0,0,0}${\bf e}(\Theta_2)$}%
}}}}
\put(6751,-3436){\makebox(0,0)[lb]{\smash{{\SetFigFont{5}{6.0}{\rmdefault}{\mddefault}{\updefault}{\color[rgb]{0,0,0}${\bf e}({\Theta_1})$}%
}}}}
\end{picture}%